# Compass-like manipulation of electronic nematicity in Sr$_3$Ru$_2$O$_7$

Masahiro Naritsuka[a,1], Izidor Benedičič[a,1], Luke C. Rhodes[a], Carolina A. Marques[a], Christopher Trainer[a], Zhiwei Li[b,2], Alexander C. Komarek[b], and Peter Wahl[a,3]

Electronic nematicity has been found in a wide range of strongly correlated electron materials, resulting in the electronic states having a symmetry that is lower than that of the crystal that hosts them. One of the most astonishing examples is Sr$_3$Ru$_2$O$_7$, in which a small in-plane component of a magnetic field induces significant resistivity anisotropy. The direction of this anisotropy follows the direction of the in-plane field. The microscopic origin of this field-induced nematicity has been a long-standing puzzle, with recent experiments suggesting a field-induced spin density wave driving the anisotropy. Here, we report spectroscopic imaging of a field-controlled anisotropy of the electronic structure at the surface of Sr$_3$Ru$_2$O$_7$. We track the electronic structure as a function of the direction of the field, revealing a continuous change with the angle. This continuous evolution suggests a mechanism based on spin–orbit coupling resulting in compass-like control of the electronic bands. The anisotropy of the electronic structure persists to temperatures about an order of magnitude higher compared to the bulk, demonstrating novel routes to stabilize such phases over a wider temperature range.

nematicity | strongly correlated electron materials | scanning tunneling microscopy | magnetism | spin–orbit coupling

## Significance

Nematicity, breaking of rotational symmetry without a reduction in translational symmetry, is found in many strongly correlated electron materials, including high temperature superconductors such as the cuprates and iron pnictides. Sr$_3$Ru$_2$O$_7$ is a particularly striking example of a nematic state, where electronic transport shows an anisotropy that depends on the direction of an applied magnetic field, however the microscopic mechanism has remained an open question. Here, using scanning tunneling microscopy, we image how the electronic states in Sr$_3$Ru$_2$O$_7$ evolve as a function of field direction. Our results demonstrate a compass-like control of the electronic structure. They can be modeled and fully understood based on the interaction of magnetism and spin–orbit coupling and their collective influence on the electronic structure.

Electronic nematicity is a symmetry reduction driven by strong correlations, typically resulting in a lowering from $C_4$ symmetry to $C_2$ symmetry, while preserving translational symmetry. Nematicity has been reported in many strongly correlated electron materials, such as the copper oxide (1) and iron-based superconductors (2, 3), heavy-fermion compounds (4), and the ruthenate Sr$_3$Ru$_2$O$_7$ (5), yet its microscopic origin remains an open question. Common to all of these materials is the observation that nematicity appears at the cusp of magnetism, close to a phase transition from a paramagnetic to a magnetically ordered ground state.

Sr$_3$Ru$_2$O$_7$ is a paramagnetic metal on the verge of magnetic order and exhibits a metamagnetic phase diagram (6) with a putative quantum critical point, which can be reached with magnetic field. Around the critical point, multiple metamagnetic transitions occur for a field along the $c$ axis near $\mu_0 H = 8$ T (6) and at temperatures below 1 K (7). It is in these phases that the resistivity exhibits in-plane anisotropy when the magnetic field is tilted by a small angle from the crystallographic $c$-direction toward one of the high-symmetry in-plane directions (Fig. 1 A and B), suggesting that a field-induced electron nematic state is realized (5). A number of mechanisms have been proposed to explain this field-induced resistivity anisotropy, for example through spin–orbit coupling (8, 9) or a Pomeranchuk instability of the Fermi surface (10). Only recently, by neutron scattering, evidence for magnetic order which aligns in field has been reported (11), possibly driven by a field-induced Lifshitz transition (12), where the resistivity anisotropy would be due to a unidirectional spin-density wave and the accompanying reconstruction of the Fermi surface. While the coupling of a spin-density wave to magnetic field can explain the field-control of the anisotropy, it is difficult to understand the continuous evolution of the resistivity as a function of the in-plane angle of the field (13). To establish a model which can link the electronic structure to the resistivity anisotropy requires detailed knowledge of the low energy electronic states in the relevant parameter regime. Here, we report symmetry breaking of the electronic states in the surface layer of Sr$_3$Ru$_2$O$_7$ that follows an applied in-plane magnetic field. From comparison with a minimal model for the electronic structure of the surface layer, we demonstrate that the field-control can be explained by in-plane ferromagnetic order and the interplay between magnetism and spin-orbit coupling. Spin–orbit coupling drives a substantial anisotropy of the electronic structure, providing a new view on electronic nematicity in Sr$_3$Ru$_2$O$_7$. Quasi-particle interference (QPI) measurements in in-plane magnetic field demonstrate a continuous evolution of the electronic structure with field direction.

Recently, quasi-particle interference at the surface of Sr$_3$Ru$_2$O$_7$ has revealed a strong breaking of $C_4$ symmetry of the electronic states (14) already in zero magnetic field.

[1]M.N. and I.B. contributed equally to this work.

[2]Present address: Key Lab for Magnetism and Magnetic Materials of the Ministry of Education, Lanzhou University, Lanzhou 730000, China.

[3]To whom correspondence may be addressed. Email: wahl@st-andrews.ac.uk.

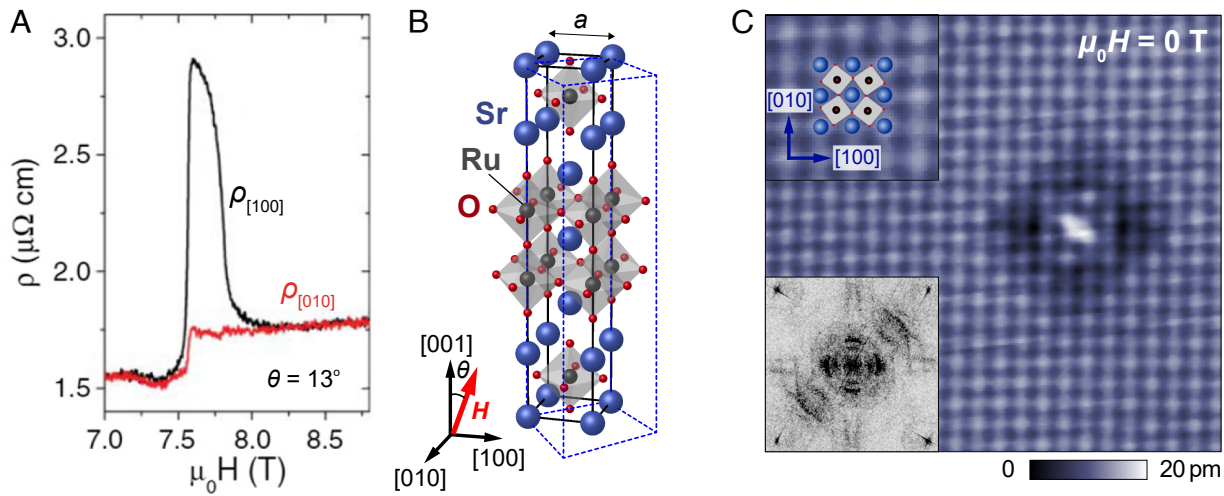

**Fig. 1.** Electronic nematicity in $Sr_3Ru_2O_7$. (A) Close to the metamagnetic critical point at $\mu_0 H \approx 7.8\,T$, applying a field with a small in-plane component $H_{ab} \parallel [100]$ (or, equivalently, [010]) induces a large anisotropy in the resistivity. Figure adapted from ref. 5. (B) Crystal structure of $Sr_3Ru_2O_7$ with the tetragonal unit cell (black lines) in side view (Blue spheres: Sr, black: Ru, red: O). Due to the octahedral rotations, the true unit cell is orthorhombic (blue dotted lines). (C) Topographic STM image showing the Sr square lattice and a twofold symmetric quasiparticle interference pattern around a defect. *Top Left* inset shows an enlarged image of the topography with the top view of the crystal structure superimposed ($V_S = 10\,mV$, $I_S = 300\,pA$). *Bottom Left* inset: Fourier transformation of a spectroscopic map acquired in zero magnetic field across similar defects in a larger area at $V = -0.3\,mV$ ($T = 80\,mK$), showing the $C_4$ symmetry breaking.

Fig. 1C shows the atomically resolved surface of $Sr_3Ru_2O_7$ in zero magnetic field and at a temperature of $T = 4.2\,K$ with an individual defect. The defect exhibits clear symmetry-breaking QPI seen as a rectangular dark shadow, and more clearly in differential conductance maps (inset of Fig. 1C). Notably, we have observed the symmetry breaking at temperatures as high as 10 K (*SI Appendix*, Fig. S7), about an order of magnitude higher than where the resistivity anisotropy is seen in the bulk, which is only observed at temperatures below 1 K (5).

The $C_4$ symmetry breaking is much more significant than one would expect just from a small orthorhombicity or anisotropy of the hopping parameters in a tight-binding description. While the bulk crystal structure is indeed orthorhombic (Fig. 1B), the difference between the lattice constants in the $a$ and $b$ directions is tiny with an orthorhombic strain of only $2(b-a)/(a+b) \sim 0.03\%$ (15), which means that often the tetragonal unit cell is used to discuss its properties (5, 11). We follow this convention and use the tetragonal unit cell for providing crystallographic directions.

In Fig. 2, we demonstrate field-control of the nematicity. Fig. 2 *A–E* show the same defect first in zero magnetic field, *A*, and then in a magnetic field of 5T applied in the *a-b*-plane, rotating the field anticlockwise from being parallel to [100], *B*, [110], *C*, [010], *D*, to parallel to [$\bar{1}$10], *E*. The symmetry-breaking QPI pattern follows the applied field, aligning with its main axis parallel to the applied field for the [110] and [$\bar{1}$10] directions. When the field is along the [100] and [010] directions, the patterns become more complex: The outline of the QPI pattern becomes almost square-shaped however still exhibiting a clear breaking of $C_4$ symmetry in the intensity distribution. On turning off the magnetic field, the QPI pattern switches back to the preferred direction observed in zero field, which is one of the [110] and [$\bar{1}$10] directions. The switching of the direction of the quasi-particle interference pattern occurs at in-plane fields larger than 0.5T (*SI Appendix*, Fig. S2), whereas the out-of-plane component of the field does not have an influence on the magnitude of the in-plane field component at which the patterns change direction. We can define a nematic order parameter $\Psi$ characterized by the angle between the [010] direction and the diagonal of a box enclosing the symmetry-breaking QPI pattern. This order parameter is zero for a $C_4$ symmetric pattern, and deviates from zero as the QPI pattern becomes $C_2$ symmetric. The magnitude of $\Psi$ corresponds to the degree of elongation of the QPI pattern, while the sign denotes its orientation. Analyzing the order parameter $\Psi$, i.e., the nematicity of the scattering pattern, as a function of field angle $\varphi$ (Fig. 2F) results in the plots shown in Fig. 2 *G* and *H*. We observe a clear nematicity of the scattering pattern for field along the [110] and [$\bar{1}$10] directions ($\varphi = 45°$ and 135°). There is a narrow range of field angles along [100] and [010], $\varphi = 0°$ and $\varphi = 90°$, where the scattering pattern becomes nearly symmetric, $\Psi \to 0°$. Apart from the change of the symmetry axis of the scattering pattern, we also observe the appearance of a checkerboard modulation dependent on the field angle (compare Fig. 2 *B* and *D*): plotting the Fourier amplitude associated with the checkerboard at $\mathbf{q}_{ckb} = (\pm 1/2, \pm 1/2)$ (Fig. 2*I*) shows significant changes with the field angle. The peak associated with the checkerboard, Fig. 2 *J* and *K*, becomes most prominent when the field is along the [100] or [010] directions ($\varphi = 0°$ and 90°, Fig. 2 *B* and *D*), whereas it becomes much weaker for field along the [110] and [$\bar{1}$10] axes ($\varphi = 45°$ and 135°, Fig. 2 *C* and *E*). The checkerboard pattern exhibits a phase shift between the two field directions, i.e., along [100] and [010], where it exhibits maximum intensity.

The strong field dependence of the quasi-particle interference patterns suggests that the surface layer is already ferromagnetic. From a tight-binding model that accounts for magnetism and spin–orbit coupling we can reproduce key experimental findings. We use a minimal model describing the electronic structure of the surface layer based on a monolayer of $Sr_2RuO_4$ with 6° octahedral rotation (described by $H_0$). Starting from a paramagnetic model, we introduce a ferromagnetic spin splitting $I$ for a magnetization direction $\mathbf{M}_0$ (16) and spin-orbit coupling $\lambda$, resulting in the Hamiltonian

$$H = H_0 + \sum_i I\mathbf{M}_0 \mathbf{S}_i + \sum_i \lambda \mathbf{L}_i \mathbf{S}_i, \qquad [1]$$

where the sums are over the two atoms in the unit cell and $\mathbf{L}_i$ and $\mathbf{S}_i$ are vectors of the angular momentum and spin operators of atom *i*. For direct comparison with the STM experiment, we calculate the continuum local density of states (cLDOS) (17–19), accounting for the tunneling matrix elements between the electronic states in the sample and the tip.

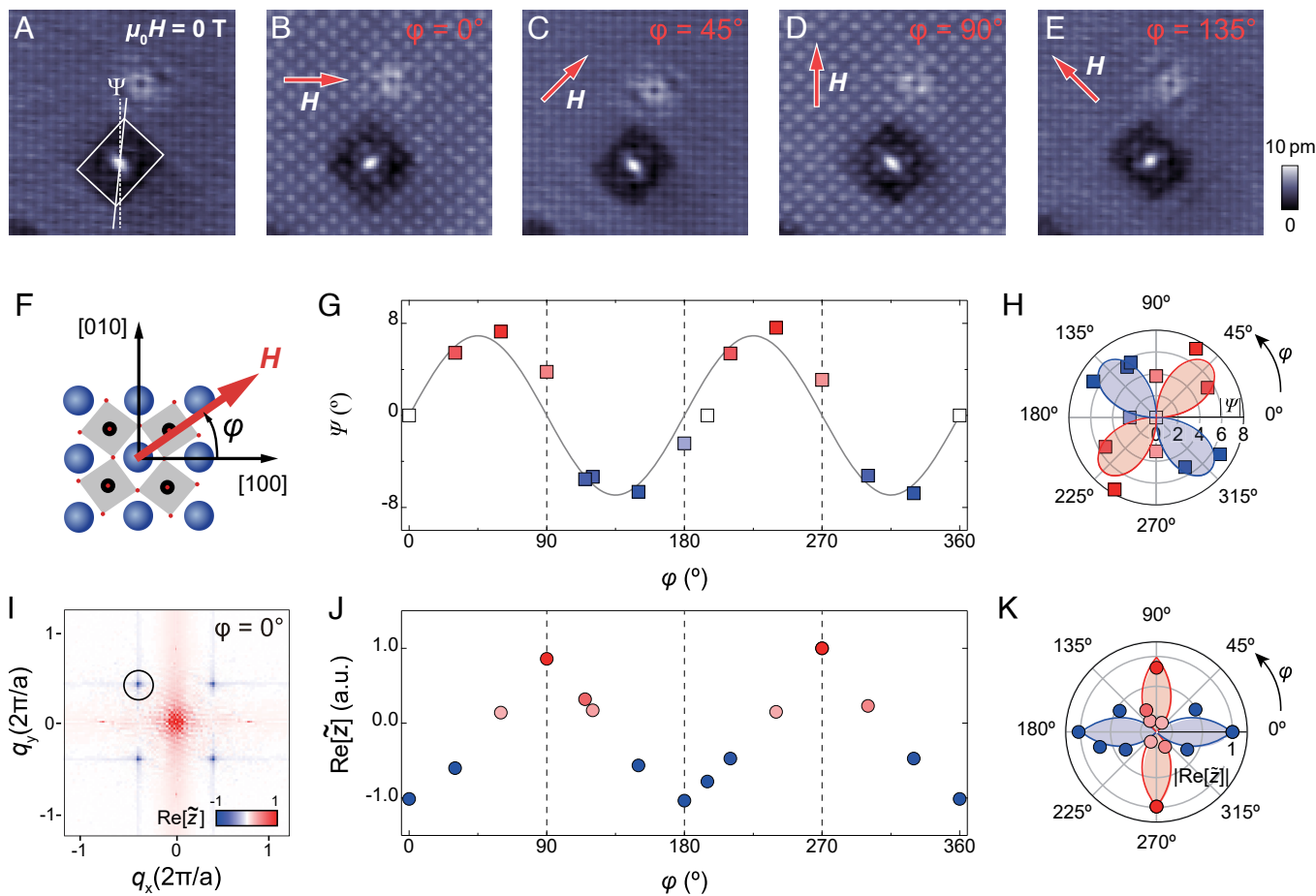

**Fig. 2.** Field-control of nematicity. (*A–E*) Topographic images around defects, *A*, at zero field with definition of the nematic order parameter Ψ, and, *B–E*, in in-plane magnetic field |μ$_0$**H**| = 5T for field directions as indicated by the red arrow ($V_S$ = 15mV, $I_S$ = 300pA, T = 4.2K). (*F*) sketch of the crystal structure and definition of angle φ in which μ$_0$**H** is applied relative to the crystallographic axes. (*G* and *H*) Dependence of nematic order parameter Ψ (extracted from the white rectangle in *A*) on the field angle φ as defined in *F*. Ψ shows sign reversal for φ = 0°, 90°, 180°, and 270°, with maxima in |Ψ| when the field is along [110] and [$\bar{1}$10], respectively (φ = 45°, 135°, 225°, and 315°). (*H*) shows the same data as in *G* in a polar plot. The solid line shows a sine function as a guide to the eye. (*I*) colour plot of the real part of the Fourier transformation Re[$\tilde{z}$(**q**)] of the topography. Red (blue) symbols indicate the same (opposite) phase relative to the topography at zero field. The Fourier peaks associated with a checkerboard modulation at **q**$_{ckb}$ = (±1/2, ±1/2) is indicated by a black circle. (*J*) the intensity of the Fourier peak at **q**$_{ckb}$ (black circle in *I*) as a function of in-plane field direction φ. The checkerboard contrast is maximal for field along [100] or [010], showing a contrast reversal between the two directions. (*K*) same data as in *J* shown in a polar plot. Lines in *G*, *H*, and *K* are drawn as guides for the eye.

In Fig. 3, we show the effect of the magnetic field and the magnetization direction for four distinct directions of the magnetization **M**$_0$. Spin–orbit coupling induces hybridization of majority- and minority-spin bands, resulting in the formation of multiple partial gaps (Fig. 3*B* and *SI Appendix*, Fig. S3 for band structure plots with orbital character). The spin–orbit coupling results in significant hybridization and thus changes the bands, which has a notable effect on the shape of the scattering patterns. We plot the cLDOS at an energy $E_1$ (Fig. 3*C*), where this band distortion is most easily visible (orange dashed line in Fig. 3*B*). With magnetization in the [100] or [010] direction (φ = 0° and 90°), bands in the Γ − S and Γ − S′ directions are equivalent (inset of Fig. 3*B*), resulting in an almost four-fold symmetric QPI pattern. In stark contrast, when the magnetization is in the [110] or [$\bar{1}$10] direction (φ = 45° and 135°), spin–orbit coupling distorts the band crossing near $E_1$. As a consequence, the cLDOS pattern shows a strong asymmetry and appears elongated in the direction of the magnetization, similar to what is seen experimentally (Fig. 2 *C* and *E*). Also close to energy $E_2$ at the upper edge of the spin-orbit induced gaps (green dashed line in Fig. 3*B*), the appearance of calculated cLDOS images is strongly dependent on the direction of the magnetization (Fig. 3*D*). This is a consequence of gaps opening in the band structure in the Γ − S and Γ − S′ directions with magnetization in the [100] or [010] directions (φ = 0° and 90°), respectively. The cLDOS at this energy shows strong checkerboard modulation with **q**$_{ckb}$ = (±1/2, ±1/2). The phase of the checkerboard changes upon rotation of the magnetization by 90°. In case of magnetization in the [110] or [$\bar{1}$10] directions, the gaps at S and S′ shift relative to each other, and the checkerboard modulation at $E_2$ is strongly suppressed. The behavior of intensity and phase of the checkerboard is consistent with what we observe in the experiment, Fig. 2 *B*, *D*, and *J*. The model also reproduces the emergence of stripe order in an out-of-plane magnetic field (20) (*SI Appendix*, Fig. S5).

Qualitatively consistent with the model, we observe distinct changes in quasi-particle interference. In Fig. 4, we show quasi-particle interference maps recorded at different in-plane directions of the applied field μ$_0$**H**. Consistent with the rotation seen in topographic images in Fig. 2 and in the model in Fig. 3, the quasi-particle inference pattern follows the magnetic field, with clear twofold anisotropic QPI patterns for field in [110] or [$\bar{1}$10] directions, whereas the QPI pattern changes when forcing the magnetization of the sample into directions of the tetragonal unit cell axes. The QPI patterns show notable changes with the field angle: The dominant symmetry breaking patterns close to $\mathbf{q} = (1/4, 1/4)$ are only seen for field along the [110] and [$\bar{1}$10] direction, which however leaves the central cloverleaf-shaped scattering pattern largely unaffected. For field along [100] or [010], the scattering vectors close to $\mathbf{q} = (1/4, 1/4)$ are suppressed, while the central pattern loses its $C_4$ symmetry, suppressing the scattering vector in the direction of the field. This is highlighted in difference maps shown in Fig. 4 *E* and *F*: Magnetic fields applied along the orthorhombic directions, [110] and [$\bar{1}$10], couple primarily to the nematic signal near $\mathbf{q} = (1/4, 1/4)$, whereas fields applied along the tetragonal axis have the dominant symmetry breaking signal along the [100] and [010] directions, the same direction in which the transport anisotropy is observed in the high-field phase in the bulk.

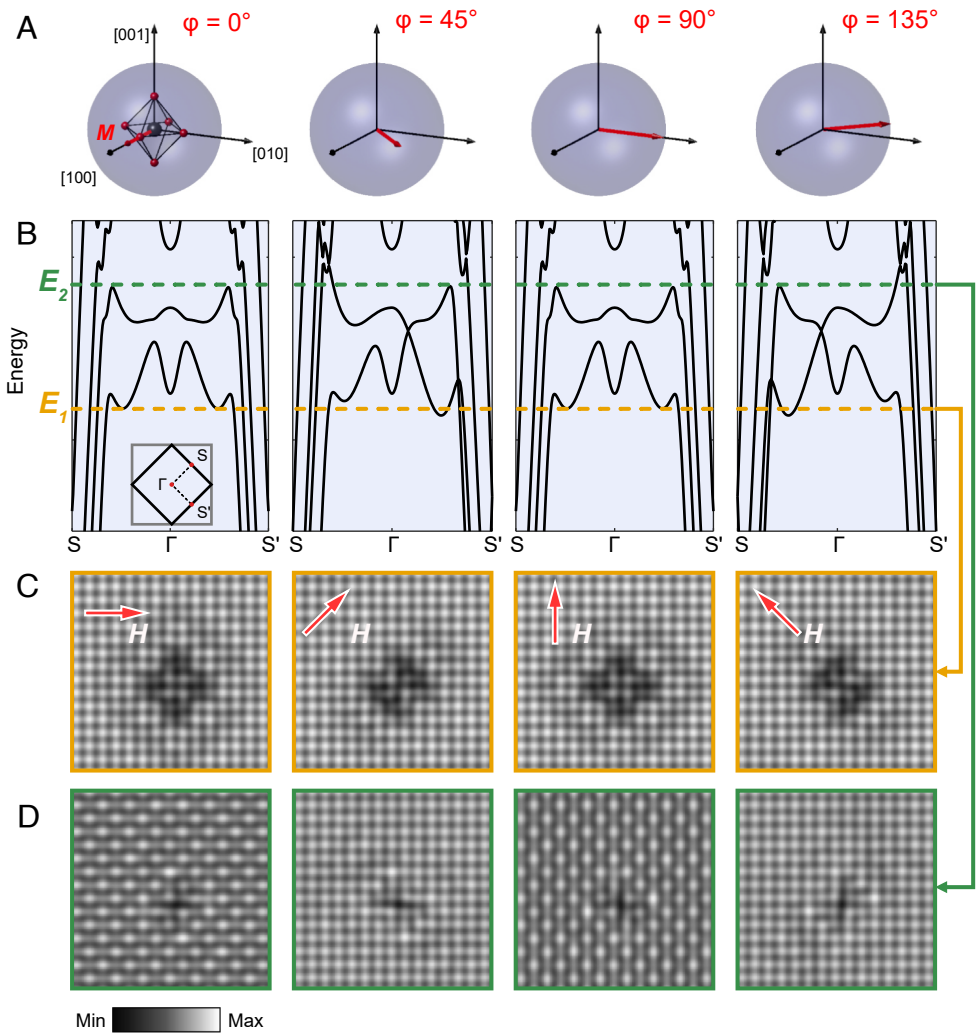

**Fig. 3.** Relation of band structure and modulated density of states. Using a tight-binding description (Eq. **1**), modeling the electronic structure of the top surface layer by a single layer of $Sr_2RuO_4$ including a magnetization **M** and spin-orbit coupling reproduces the key findings of Fig. 2. To model the influence of the magnetic field on the electronic structure, we assume that the magnetization **M** points in the direction of the applied field (red arrow), (*A*). As a consequence of spin–orbit coupling, the band structure obtained from the minimal model (for details see main text and *Materials and Methods*), (*B*), depends on the in-plane direction of the magnetization. For direct comparison with the experimental data, we calculate the real-space continuum local density of states (cLDOS), (*C*), for the in-plane magnetisation directions shown in *A* at energy $E_1$ indicated by the orange horizontal line in *B*, showing symmetry breaking QPI patterns, with the symmetry breaking most obvious for $\varphi = 45°$ and $135°$, as also seen in Fig. 2 *C* and *E*. (*D*) cLDOS for magnetization directions indicated in *A* at energy $E_2$, green line in *B*, showing the appearance of a checkerboard for $\varphi = 0°$ and $90°$ as in Fig. 2 *B* and *D*.

Our results demonstrate an electronic nematicity that can be controlled in a compass-like fashion, where the direction of the nematicity is dictated by the direction of an external magnetic field.

Our results, and their description in our minimal model, suggest that the surface layer of $Sr_3Ru_2O_7$ becomes magnetic, possibly due to the surface relaxation and interface with the vacuum, stabilizing a similar magnetically ordered phase as found at high fields in the bulk. From LEED-IV measurements (21) and density functional theory (DFT) calculations of surface slabs (22), the octahedral rotation in the surface layer is $\sim 12°$ (21), significantly larger compared to the bulk (8°). Both, from experiments (23) and calculations (24–27), it is found that increased octahedral rotations in ruthenates tend to stabilize magnetism. For the bulk, DFT suggests a magnetic ground state (28) even in zero magnetic field. Combined with spin-orbit coupling, the surface magnetism can explain the nematicity as demonstrated by our minimal model. We use a value of spin–orbit coupling obtained for Ru atoms (29), but note that there is evidence that

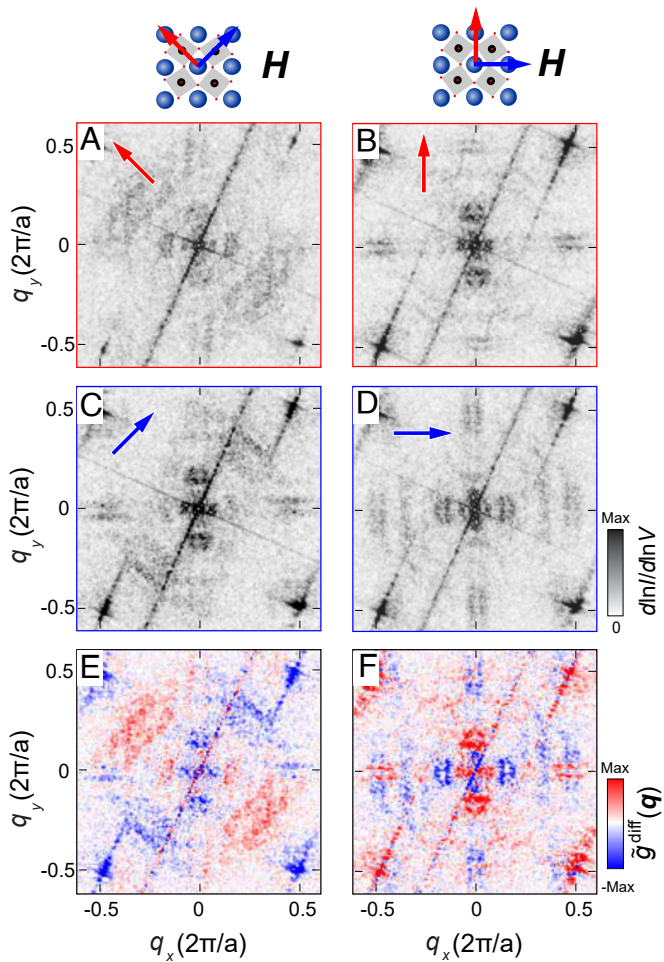

**Fig. 4.** Quasiparticle interference imaging. (*A*–*D*) Fourier transform of d ln *I*/d ln *V* maps at different directions of the in-plane field **H** along [$\bar{1}$10] (*A*), [010] (*B*), [110] (*C*) and [100] (*D*), shown at $V = -0.8$ mV ($V_S = 10$ mV, $I_S = 100$ pA, $T = 4.2$ K, $\mu_0 H = 5$ T, in-plane field direction is indicated by an arrow in each panel). (*E*) Difference of Fourier transforms of spectroscopic maps in *A* and *C* with magnetic field $\mu_0 H = 5$ T applied in the orthorhombic directions (**H** ∥ [$\bar{1}$10] and **H** ∥ [110]). (*F*) Same as *E* but for *B* and *D* with magnetic fields applied in the tetragonal directions (**H** ∥ [010] and **H** ∥ [100]).

correlations result in an increased spin–orbit coupling compared to simple renormalization of the DFT value (30). The model qualitatively reproduces the experimentally observed behavior and shows how the combination of ferromagnetism and large spin-orbit coupling results in a significant anisotropy of the electronic structure, responsible for sensitivity to the magnetic field direction—a similar mechanism has been proposed to describe the resistivity anisotropy in the bulk (8). Without spin-orbit coupling, the change in magnetization direction does not result in changes of the electronic structure (*SI Appendix*, Fig. S4). We here observe the nematicity already in zero magnetic field and at temperatures as high as 15 K, whereas in the bulk the transport anisotropy is only observed below 1 K and in high fields (5). From measurements in vector magnetic fields, we can map out the magnetic anisotropy of the system, showing that the surface layer has an easy plane anisotropy in the plane of the surface, with the preferred directions of the magnetization in the orthorhombic *a* and *b* directions ([110] and [$\bar{1}$10]). Due to the orthorhombicity, these directions are not identical, confirmed by a preferred direction of the symmetry breaking QPI patterns in zero field, as well as for the bulk by an anisotropy in the magnetization

(*SI Appendix*, Fig. S1). The data shown here provide a complete picture of field-controlled nematicity of the surface layer. They exhibit intriguing analogies with the bulk properties, suggesting that a full description of the field-induced transport anisotropy has to account for spin–orbit coupling and how it modifies the Fermi surface in a magnetic ground state. The results suggest an alternative interpretation of the nematic phase detected in the bulk. The continuous evolution of the electronic structure with the field angle that we observe here is expected to result in a continuous dependence of the resistivity on the field angle as observed experimentally (13). The results might also be consistent with the low-moment order detected in neutron scattering (11), if the field-induced order is a consequence of an anisotropic spin susceptibility due to the anisotropy of the electronic structure. We note that the central scattering pattern observed in our QPI indeed adopts a similar anisotropy as observed in neutron scattering. For field along [010] and [100], the scattering pattern exhibits twofold symmetry with the strongest intensity near **q** ~ $(0, \pm 1/4)$ for field along [010] (Fig. 4*B*), and in a 90° rotated direction for field along [100] (Fig. 4*D* and *SI Appendix*, Fig. S6). The patterns here are not due to static order though, but due to the electronic susceptibility resulting in quasi-particle interference and Friedel-like oscillations around defects.

There are important open questions on the relation of our results to the bulk properties of $Sr_3Ru_2O_7$: The nematicity in the transport properties (5) as well as the magnetic order detected in neutron scattering (11) show the largest anisotropy between the tetragonal axis, i.e., [100] and [010], whereas from our data here as well as magnetization, the anisotropy is strongest between the orthorhombic axis, i.e., the [110] and [$\bar{1}$10] directions. Also, in the bulk a significant out-of-plane field is required to drive the material into a polarized state in the first place.

It is remarkable that such small magnetic fields, below 1 T, have such a dramatic effect on the electronic structure. The energy scale associated with fields is only $g\mu_B\mu_0 H \sim 0.1$ meV, much smaller than the energy scale of thermally excited electrons at the temperature of the experiment.

Our results demonstrate a mechanism of field-control of the electronic structure based on the interplay of spin–orbit coupling and magnetism. We see clear signatures of the spin–orbit coupling rendering magnetic and electronic properties anisotropic, underlining the importance of the interplay between structural, magnetic and electronic degrees of freedom in the ruthenates. Due to the strong spin–orbit coupling, a compass-like control of the electronic structure is realized, whose anisotropy follows the direction of an externally applied magnetic field. Our results demonstrate a close link between nematicity at the microscopic scale and magnetism in the surface layer, providing important new insights for our understanding of the resistivity anisotropy in $Sr_3Ru_2O_7$ with potentially far-reaching implications for a wide range of materials. Similar physics is expected in any ferromagnetic material with strong spin–orbit coupling. Our minimal model provides direct insight into how materials can be optimized for field-controlled transport anisotropies, opening the door to tune them for possible applications.

## Materials and Methods

**Single Crystal Growth.** Single crystals of $Sr_3Ru_2O_7$ have been grown in an image furnace using the method described in ref. 14. The crystals have been characterized by low temperature transport measurements (see *SI Appendix*, Text S1 for details) to verify their quality and that they exhibit the behavior found in high-purity crystals (5).

**Scanning Tunneling Microscopy.** Low temperature scanning tunneling microscopy (STM) measurements have been performed in a home-built STM with a sapphire head (31) mounted in a 9/5 T vector magnet (32). Additional supporting measurements were performed in an STM mounted in a dilution refrigerator (33). For all measurements, samples were cleaved in-situ at low temperature (≈20 K) before inserting them into the STM. Measurements shown here were performed at 4.2 K unless stated otherwise. Tunneling spectra were acquired applying the bias voltage to the sample and using the standard lock-in technique.

**Minimal Model.** To model the appearance of the STM images of the surface of $Sr_3Ru_2O_7$, we use a minimal model based on a free-standing monolayer of $Sr_2RuO_4$ to calculate the local density of states (LDOS). While we show topographic images, e.g., in Fig. 2, the anisotropy seen in these images is due to an anisotropy in the electronic structure, so will be apparent in simulated images of the differential conductance. The model includes an octahedral rotation of $6°$, similar to that of $Sr_3Ru_2O_7$.

We found that this model, which neglects the bilayer splitting of $Sr_3Ru_2O_7$, captures the qualitative physical phenomena that we observe in our measurements once spin–orbit coupling and in-plane ferromagnetism are included.

We obtain the tight-binding model for a free-standing $Sr_2RuO_4$ layer with octahedral rotations from a paramagnetic DFT calculation using Quantum Espresso (34), using the Perdew–Burke–Ernzerhof exchange correlation functional, a **k**-grid of $8 \times 8 \times 1$, a wavefunction cutoff of $E_{cut,wfc} = 90$Ry, and a density cutoff of $E_{cut,\rho} = 720$Ry. The ground state wavefunction of this model is projected onto a tight-binding model consisting of the $t_{2g}$ orbitals to obtain $H_0$ using Wannier90 (35). We then add the exchange splitting $I = 400$ meV and the spin-orbit coupling $\lambda = 173$ meV to the tight-binding model as described in the main text. We do not account for renormalization or adjust the chemical potential to ensure charge conservation, as the purpose of the minimal model is to demonstrate which are the minimal ingredients required to reproduce the experimentally observed phenomena. We expect that a full model will require accounting for these as well as the additional bands due to the bilayer structure. Relative to the chemical potential of the paramagnetic DFT bandstructure, energies $E_1$ and $E_2$ correspond to −282 meV and −215 meV, respectively.

For comparison with the STM data, we perform continuum local density of states (cLDOS) calculations, modeling the QPI patterns using the usual $T$-matrix approach. We calculate the unperturbed Green's function from

$$G_{0,\sigma}(\mathbf{k}, \epsilon) = \sum_n \frac{\xi_{n\sigma}^\dagger(\mathbf{k})\xi_{n\sigma}(\mathbf{k})}{\omega - E_{n\sigma}(\mathbf{k}) + i\eta}. \quad [2]$$

From the unperturbed Green's function, we obtain the Green's function in presence of a scatterer from

$$G_\sigma(\mathbf{R}, \mathbf{R}', \omega) = G_{0,\sigma}(\mathbf{R} - \mathbf{R}', \omega) + G_{0,\sigma}(\mathbf{R}, \omega)T_\sigma(\omega)G_{0,\sigma}(\mathbf{R}', \omega), \quad [3]$$

using the $T$-matrix

$$T_\sigma = \frac{V_\sigma}{1 - V_\sigma G_{0,\sigma}(0, \omega)}, \quad [4]$$

where $V_\sigma$ is the scattering potential. Here we use $V_\sigma = 1$eV. To simulate spatial maps of the differential conductance $g(\mathbf{r}, V)$, we carry out the transformation to the continuum Green's function (17–19) through

$$G_\sigma(\mathbf{r}, \mathbf{r}', \omega) = \sum_{\mathbf{R}, \mathbf{R}', \mu, \nu} G_\sigma^{\mu,\nu}(\mathbf{R}, \mathbf{R}', \omega) w_{\mathbf{R},\mu}(\mathbf{r}) w_{\mathbf{R}',\nu}(\mathbf{r}'), \quad [5]$$

where the wave functions $w_{\mathbf{R},\mu}(\mathbf{r})$ are obtained from Wannier90, using a modified version that preserves the relative sign of the wave functions. The local density of states $\rho(\mathbf{r}, \omega)$ is then obtained from

$$\rho(\mathbf{r}, \omega) = -\sum_\sigma \frac{1}{\pi} \text{Im} G_\sigma(\mathbf{r}, \mathbf{r}, \omega). \quad [6]$$

For the calculations shown in Fig. 3, we have performed continuum QPI calculations on a **k**-grid with $1,024 \times 1,024$ points and a real space lattice of $16 \times 16$ unit cells, using an energy broadening $\eta = 2$ meV.

**Data, Materials, and Software Availability.** Data underpinning the manuscript will be deposited in a public repository as text data. Data have been deposited in Research repository St Andrews (https://doi.org/10.17630/38c4e35c-5274-4e9b-a4ab-c90c96d1e37f).

**ACKNOWLEDGMENTS.** We gratefully acknowledge discussions with Stephen Hayden, Chris Hooley, Peter Littlewood and Andy Mackenzie. M.N., C.A.M., and P.W. acknowledge funding from EPSRC through EP/R031924/1 and I.B. through the International Max Planck Research School for Chemistry and Physics of Quantum Materials. L.C.R. was supported through a fellowship from the Royal Commission for the Exhibition of 1851. C.A.M. further acknowledges funding from EPSRC through EP/L015110/1.

Author affiliations: [a]Scottish Universities Physics Alliance, School of Physics and Astronomy, University of St Andrews, North Haugh, St Andrews, KY16 9SS, United Kingdom; and [b]Max Planck Institute for Chemical Physics of Solids, Dresden 01187, Germany

# Supplementary material for 'Compass-like manipulation of electronic nematicity in $Sr_3Ru_2O_7$'

Masahiro Naritsuka,[1,*] Izidor Benedičič,[1,*] Luke C. Rhodes,[1] Carolina A. Marques,[1] Christopher Trainer,[1] Zhiwei Li,[2,†] Alexander C. Komarek,[2] and Peter Wahl[1,‡]

[1]SUPA, School of Physics and Astronomy,
University of St Andrews, North Haugh,
St Andrews, Fife, KY16 9SS, United Kingdom

[2]Max Planck Institute for Chemical Physics of Solids,
Nöthnitzer Straße 40, 01187 Dresden, Germany

---

[*]These authors contributed equally.

[†]Current address: Key Lab for Magnetism and Magnetic Materials of the Ministry of Education, Lanzhou University, Lanzhou 730000, China

[‡]Correspondence to: wahl@st-andrews.ac.uk

## S1. CHARACTERIZATION OF SAMPLE

Single crystal samples of $Sr_3Ru_2O_7$ used for the experiments here were from the same batch as the ones used in Ref. [1]. Samples were characterized by resistivity (RRR $\sim$ 102) and specific heat measurements as well as Transmission Electron Microscopy. These results are available in section S1 of the supplementary material of Ref. 1.

### A. Magnetization measurements

The in-plane field dependence of the metamagnetic transitions of $Sr_3Ru_2O_7$ were characterized by magnetization measurements. The measurements were carried out using a Quantum Design MPMS3 SQUID magnetometer capable of applying a magnetic field of up to 7T and equipped with a sample rotator. The sample was mounted on the rotator such that the field could be applied in any direction in the sample $a - b$ plane. The samples were cut along the edges of the tetragonal unit cell ([1 0 0] and [0 1 0] directions in the orthorhombic unit cell). The alignment was checked using X-ray diffraction. When mounting the sample for measurements, the edges were aligned with the straight edges of the sample stage of the rotator, thus allowing the field direction for the measurement to be determined relative to the crystallographic directions. To account for the diamagnetic background of the sample rotator, the measurements presented in the following were repeated with empty rotator only, allowing for subtraction of the background signal from the rotator.

The measurements in the MPMS were conducted at 1.8K. The sample was rotated through a 180° angle in 6.4° steps. For each 6.4° increment the field was ramped from 4.5T to 7T, the field range of the metamagnetic transitions in $Sr_3RuO_7$. The derivative of the magnetization data, i.e. the magnetic susceptibility of the sample, is plotted as a color plot in Fig. S1a. The magnetization data for fields applied along the different crystallographic high-symmetry directions is shown in Fig. S1b. The resulting susceptibility curves are shown in the inset of Fig. S1b. The two metamagnetic transitions appear as peaks in the magnetic susceptibility, the first as a sharp peak at $\sim 5$T and the second as a broader peak slightly above 6T. We determine the field at which the metamagnetic transition occurs by fitting a Gaussian function to these peaks.

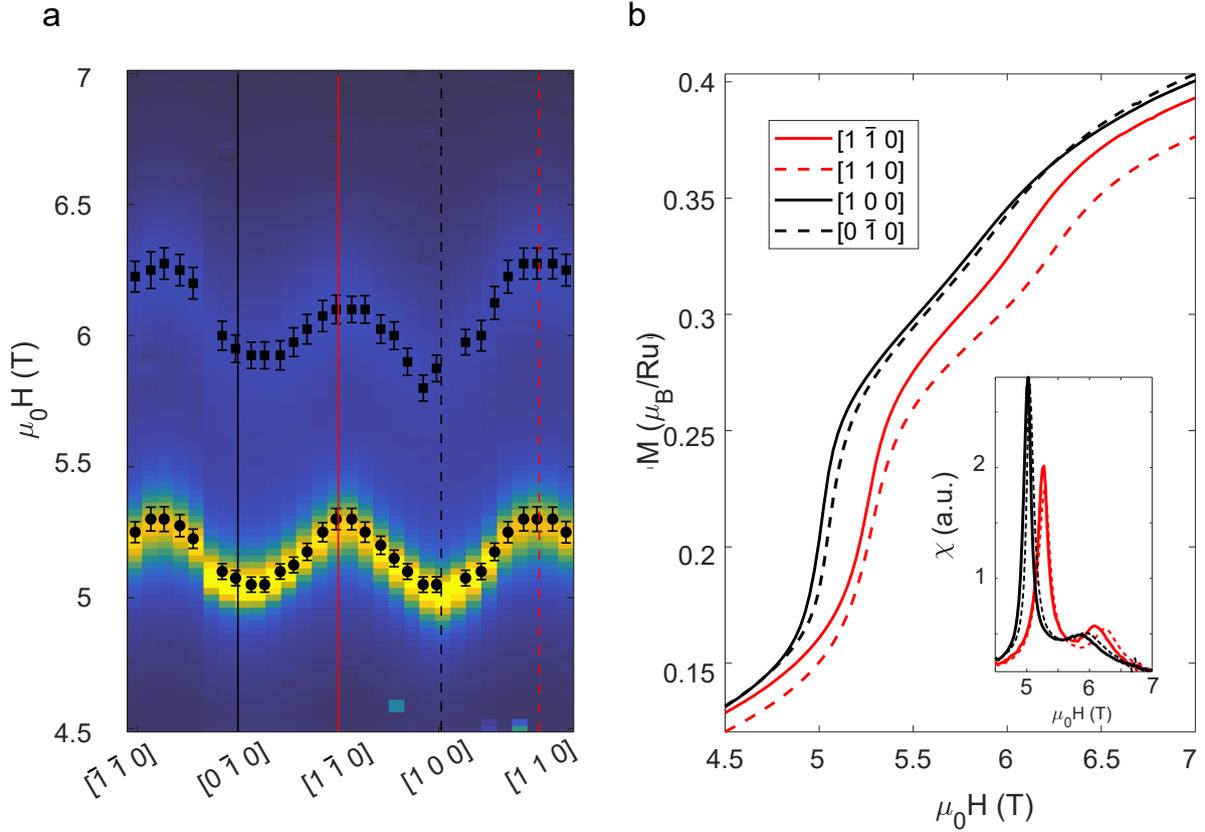

FIG. S1: **Angle dependence of in-plane magnetization.** (a) Color plot of the magnetic susceptibility of $Sr_3Ru_2O_7$ as a function of field angle in the sample $a-b$ plane, the fields at which the metamagnetic transitions occur are plotted as points on top of the color map plot. The metamagnetic transitions and their associated errors are determined by fitting a Gaussian function to the peaks in the magnetic susceptibility. (b) Plot of the magnetization for fields applied along the orthorhombic ($[1\,1\,0]$, $[1\,\bar{1}\,0]$) and tetragonal ($[1\,0\,0]$, $[0\,\bar{1}\,0]$) axes. The inset shows the corresponding magnetic susceptibility curves.

From the results of the magnetization measurement, we find that the magnitude of the field required to induce the metamagnetic transition depends on the orientation of the field in the crystallographic $a-b$ plane. The first transition at about 5T exhibits a four-fold symmetry with respect to the field direction. The transition occurs at higher fields for field along the orthorhombic $[1\,1\,0]$ and $[1\,\bar{1}\,0]$ directions, and at lower field along the tetragonal axes ($[1\,0\,0]$ and $[0\,1\,0]$). The transition at higher fields (around 6T) exhibits only a two-fold symmetry as a function of field angle. The maximum field required to induce this transition is for fields along the orthorhombic axes ($[1\,1\,0]$ and $[1\,\bar{1}\,0]$), but for one of them, the transition occurs for a lower field than for the other.

## S2. DATA PROCESSING

### A. Determination of nematic order parameter $\Psi$

The following assumptions are made in defining the nematic order parameter $\Psi$: 1) the shape of the scattering pattern around the defect is rectangular; 2) the directions of the edges of the rectangle are always along the crystallographic $a-$ and $b-$axis; 3) $a > b$ at zero field. $\Psi$ is defined as,
$$\Psi = \arctan\frac{a}{b} - 45°,$$
so that $\Psi$ becomes zero when scattering from the defect becomes four-fold symmetric.

The dependence of the order parameter on the magnetic field strength when the magnetic field is applied in the orthorhombic directions ($[1\,1\,0]$ and $[1\,\bar{1}\,0]$) is shown in Fig. S2. The magnetic field strength at which the order parameter becomes zero is not strongly affected by the magnetic field strength in the $c$-axis direction.

### B. Phase-referenced Fourier transformation

In Fig. 2j, we show that the intensity and relative phase of the checkerboard charge order varies periodically with the direction of the applied in-plane magnetic field. The surface layer of $Sr_3Ru_2O_7$ has two inequivalent but symmetry-related Sr sites due to the rotation of the $RuO_6$ octahedra around the $c$-axis. This is reflected in the checkerboard pattern of the brighter and darker Sr atoms in topographic images. When the field direction is set to $\varphi = 0°$, Sr atoms at A sites are brighter than ones at B sites. As $\varphi$ increases, the difference in brightness between the two sites becomes smaller toward $\varphi = 45°$. With further rotation from $\varphi = 45°$ to $\varphi = 90°$, the B sites become brighter than the A sites. The reveral of the contrast between the two sublattices can be clearly seen from Fig. 2b, d.

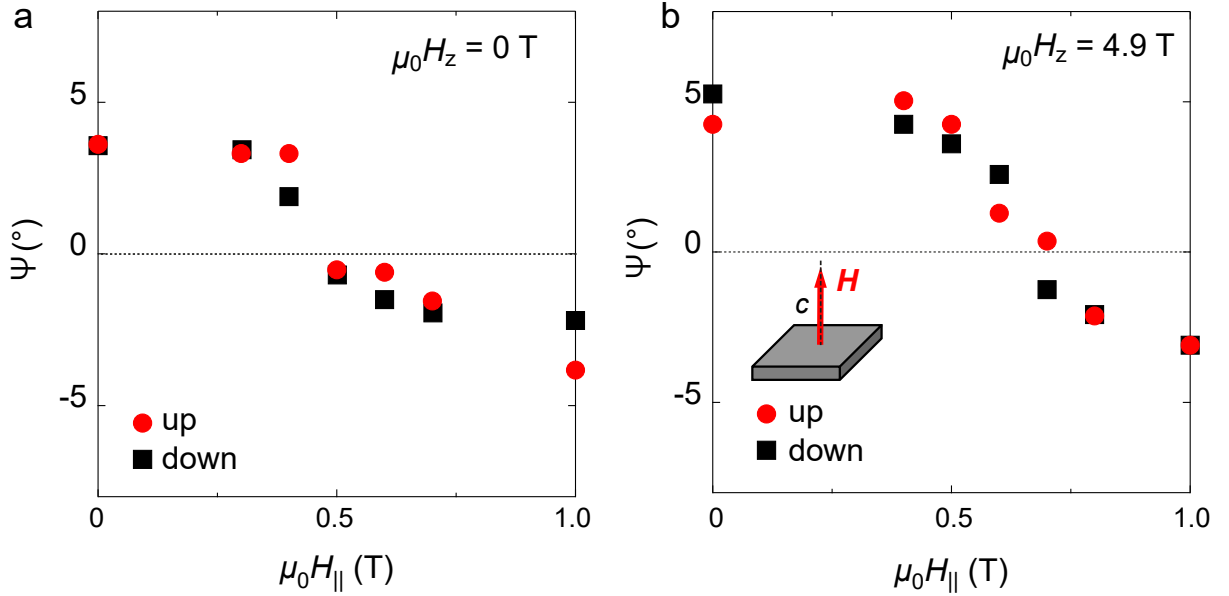

FIG. S2: **Nematic order parameter Ψ as a function of the in-plane field $\mu_0 H_\parallel$.** (a, b) Order parameter Ψ as a function of in-plane field for an out-of-plane component of the field of (a) $\mu_0 H_z = 0$T and (b) $\mu_0 H_z = 4.9$T. The in-plane field is applied in the in-plane direction normal to the preferred direction of the nematicity in zero field ($\varphi = 45°$). The field at which the nematicity switches direction, i.e. $\Psi \to 0$, is $\sim 0.5$T for $\mu_0 H_z = 0$T (a) and $\sim 0.6$T for $\mu_0 H_z = 4.9$T (b). The upsweep (red circles) and downsweep (black rectangles) of the magnetic field are almost identical, showing, within the error of our measurements, no indication of hysteresis.

In order to capture this behavior of the checkerboard charge order quantitatively, we use a phase-referenced Fourier transformation of topographic images. The checkerboard charge order appears as two pairs of Bragg peaks at $\mathbf{q}_{\text{ckb}} = (\pm 1/2, \pm 1/2)$. The intensity of the checkerboard appears in the intensity of the Fourier peak at $\mathbf{q}_{\text{ckb}}$. The relative phase change is reflected in the information of the phase of the Fourier peaks. The phase is represented by $\theta$ in the following,

$$\theta(\mathbf{q}_{\text{ckb}}, \varphi) = \arccos\left(\text{Re}\left[\frac{\tilde{z}(\mathbf{q}_{\text{ckb}}, \varphi)}{|\tilde{z}(\mathbf{q}_{\text{ckb}}, \varphi)|}\right]\right),$$

where $\tilde{z}(\mathbf{q}, \varphi)$ is the Fourier transform of the topography obtained at field angle $\varphi$. The choice of the origin of the phase is arbitrary, so for simplicity, the phase $\theta$ is referenced relative to $\varphi = 0°$. As shown in Fig. 2j, $\cos(\theta(\mathbf{q}_{\text{ckb}}, \varphi)) = -1$ at $\varphi = 0°$, 180° and 360°, and $\cos(\theta(\mathbf{q}_{\text{ckb}}, \varphi)) = 1$ at $\varphi = 90°$, 270°. Four sign reversals are observed around $\varphi = 45°$, 135°, 225°, 315° while the magnetic field is rotated by 360°.

To be able to compare the phase between consecutive topographic images, we apply the Lawler algorithm[2] to correct for any drift and cropped the exact same region around a set of defects from topographic images, ensuring that the global phase is the same across the whole data set.

## C. Processing of QPI data

All spectroscopic maps acquired for the four different in-plane directions of the magnetic field shown in Fig. 4a-d were performed at the same location of the sample, using the same tip and the same measurement conditions (set point, temperature), changing only the magnetic field direction. The resulting maps were cropped to the exact same region, so that the Fourier transformations show quasi-particle interference originating from the same atomic-scale area of the sample. To correct for minor drift, the atomic peaks were mapped to be exactly on a square with same lattice constant. Finally, a rotation operation was performed so that the Sr-Sr direction is aligned with the vertical and horizontal directions of the image. No symmetrization was performed on any of the QPI data. The resulting dI/dV map is shown in Fig. S5. In the main text, we show $\text{dln}I/\text{dln}V(V) = \frac{(\text{d}I/\text{d}V)(V)}{I(V)/V}$ to remove the influence from the set-point effect.

## S3. TIGHT-BINDING MODELS

### A. Band structure plots with orbital character

In addition to band structure plots in Fig. 3 in the main text, we show in Fig. S3 band structures plotted on high-symmetry paths of a one-atom unit cell. The orbital characters are indicated by the colour of the bands.

To demonstrate that our minimal tight-binding model can reproduce the main features of the low-energy electronic structure, in Fig. S4 we calculate the cLDOS with magnetisation tilted out of the *ab*-plane and compare it with the case with magnetisation completely lying within the plane. With tilted magnetisation, we observe emergence of stripes connecting the Sr atoms in $[\bar{1}\,1\,0]$ direction. This is qualitatively consistent with previous experimental reports [1] where the emergence of stripe order with out-of-plane magnetic field was observed, providing further evidence for the validity of our model.

---

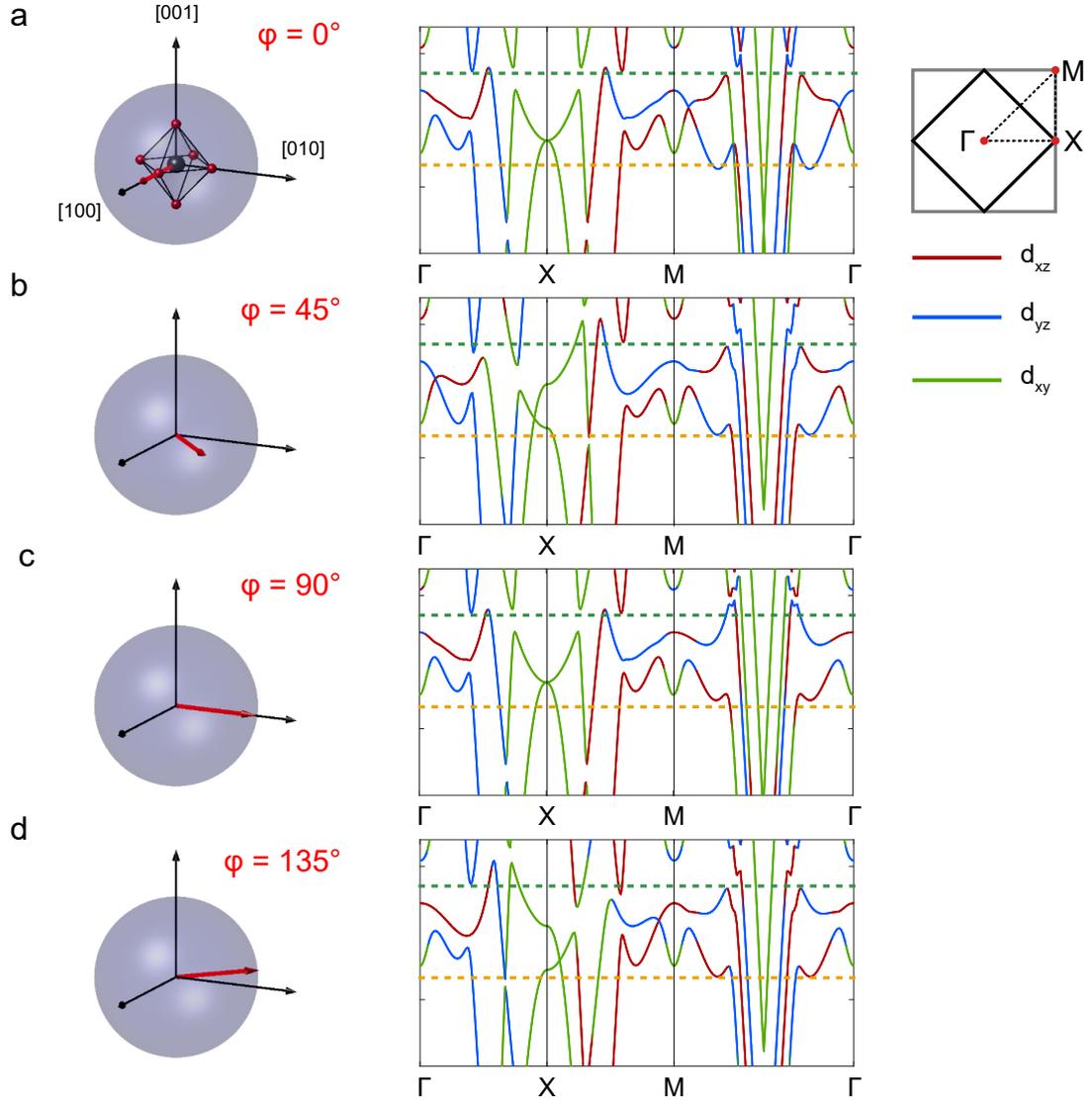

FIG. S3: **Orbitally-resolved band structure with different magnetisation directions a-d**, Band structure plotted in whole one-atom unit cell Brillouin zone. Bands color at each k-point is determined by its maximum orbital character.

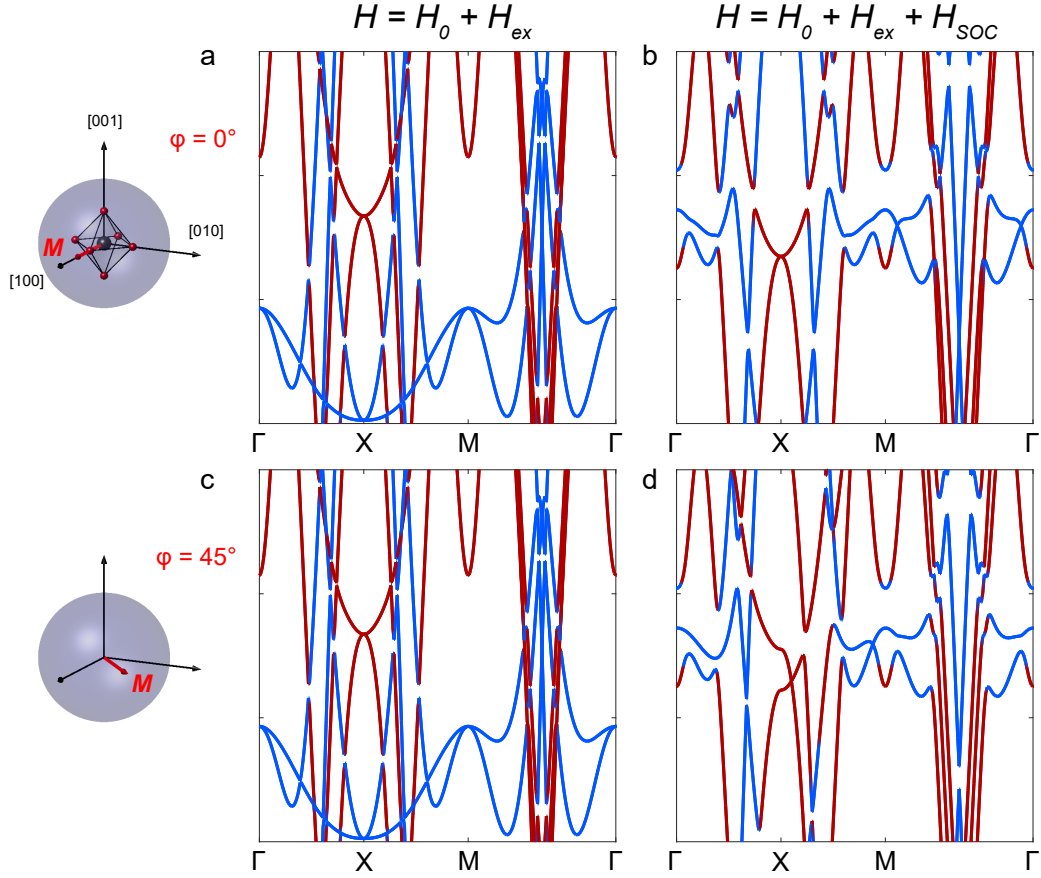

FIG. S4: **Effects of magnetisation and spin-orbit interaction a**, Band structure plotted in whole one-atom unit cell Brillouin zone, with $\vec{M} \parallel [1\,0\,0]$ but without spin-orbit coupling. Red and blue colours denote majority and minority spin character, respectively. **b**, Band structure with $\vec{M} \parallel [1\,0\,0]$ and spin-orbit interaction included in the calculation. **c**, Band structure with $\vec{M} \parallel [1\,1\,0]$ without spin-orbit coupling. **d**, Band structure with $\vec{M} \parallel [1\,1\,0]$ and spin-orbit interaction included in the calculation.

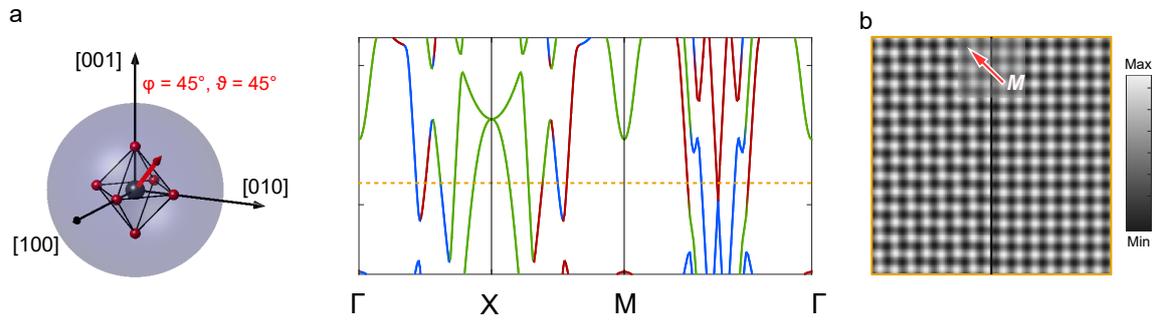

FIG. S5: **Electronic structure with out-of-plane tilting a**, Band structure of a minimal model with magnetisation tilted by 45° from the $ab$ plane. **b**, cLDOS calculated at energy indicated by orange dashed line in **a**. Left: cLDOS of a model with magnetisation tilted by $\vartheta = 45°$ from the $ab$ plane. Right: cLDOS of a model with magnetisation lying entirely within the $ab$ plane, $\theta = 0°$.

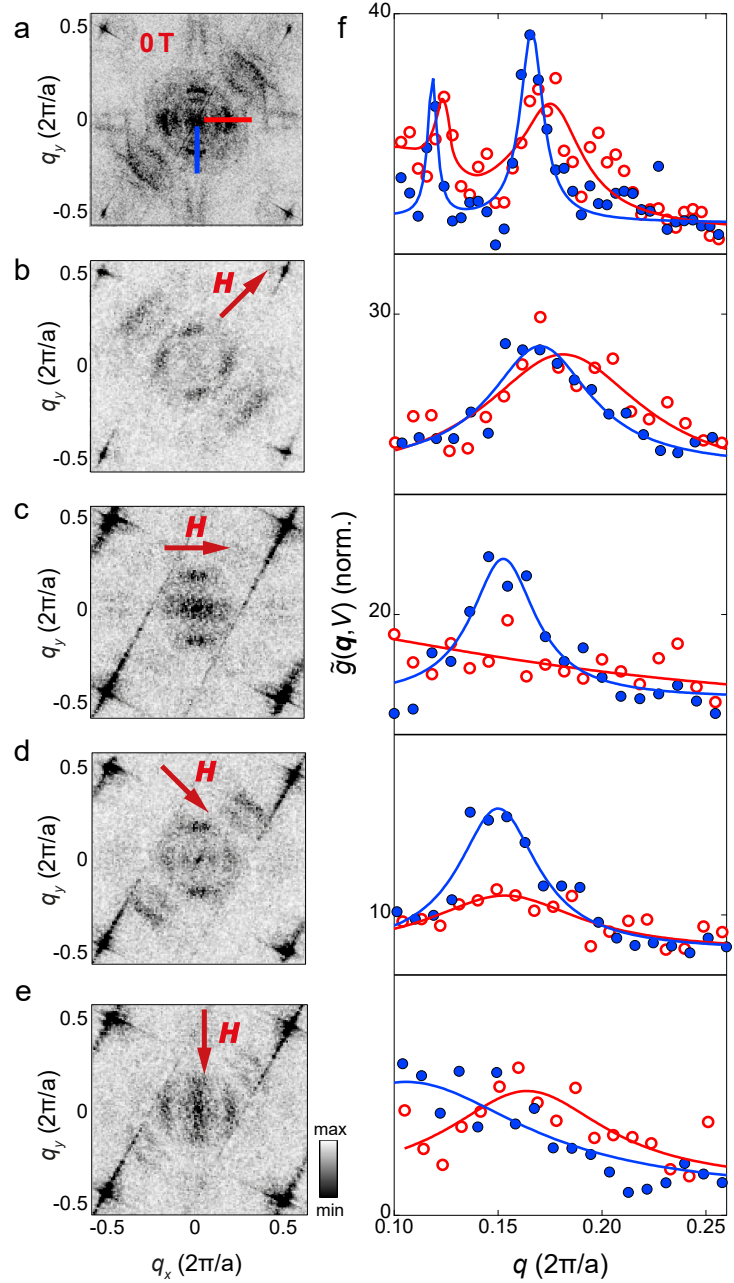

FIG. S6: **Quasiparticle interference imaging. a-e.** Fourier transform of spectroscopic map, shown at **a.** $V = -0.3$mV ($V_s = 8$mV, $I_s = 700$pA, $V_L = 600$ $\mu$V, $T = 80$mK, $H = 0$) and **b-e**. $V = 0$mV ($V_s = 10$mV, $I_s = 100$pA, $T = 4.2$K, $\mu_0 H = 5$T.) **f.** Line cuts of **a-e**. along [1 0 0] ($q_x$, red) and [0 1 0] ($q_y$, blue) directions. The lines are fitted with Lorentzians as a guide for the eye.

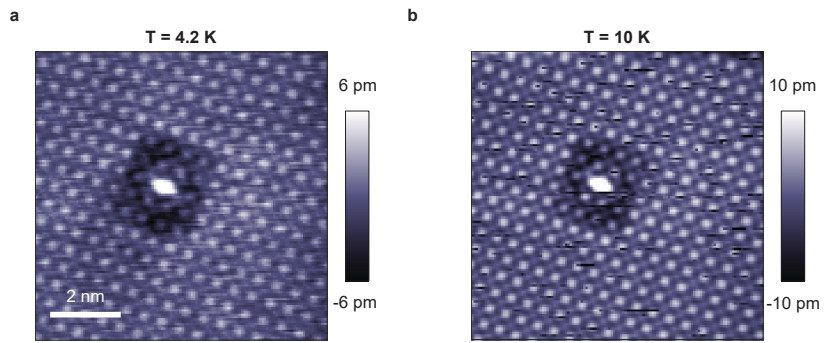

FIG. S7: **Temperature dependence of symmetry breaking a**. Topography of a single defect at $T = 4.2$K. **b**. Topography of the same defect at $T = 10$K.